\documentclass[a4paper]{jpconf}
\usepackage{graphicx}

\newcommand{\ket}{\rangle }
\newcommand{\bra}{\langle }

\newcommand{\ve}{\varepsilon}
\newcommand{\dket}{\ket\ket}
\newcommand{\dbra}{\bra\bra}

\newcommand{\Vect}[1]{\mbox{\boldmath$#1$}}

\begin{document}
\title{Photovoltaic Berry curvature in the honeycomb lattice}

\author{Takashi Oka$^{1,2}$ and Hideo Aoki$^1$}

\address{$^{1}$Department of Physics, University of Tokyo, Hongo, Tokyo 113-0033, Japan,\\
$^{2}$Theoretische Physik, ETH Zurich, 8093 Zurich, Switzerland}
\ead{oka@cms.phys.s.u-tokyo.ac.jp}

\begin{abstract} 
Photovoltaic Hall effect --- the Hall effect
induced by intense, circularly-polarized light in the absence of 
static magnetic fields --- has been 
proposed in Phys. Rev. {\bf B}  79, 081406R (2009) 
for graphene where a massless Dirac dispersion is realized. 
The photovoltaic Berry curvature (a nonequilibrium 
extension of the standard Berry curvature) is the key quantity to 
understand this effect, which appears in the
Kubo formula extended to Hall transport in the presence of
strong AC field backgrounds.  Here we elaborate the properties of 
the photovoltaic curvature such as the frequency
and field strength dependence in the honeycomb lattice.
\end{abstract}

\section{Introduction}
Non-linear effects in electronic systems 
are an interesting playground to look for novel 
transport properties that cannot occur in equilibrium. 
A famous example is the Franz-Keldysh effect in semiconductors, 
where the band edge exhibits red-shifts 
in static or AC electric fields. 
This can be considered as an effect of the 
deformation of the electron wavefunction, i.e., leakage to the gap region,
in strong electric fields, 
which leads to the change in response properties. 

\begin{figure}[t]
\centering 
\includegraphics[width=6cm]{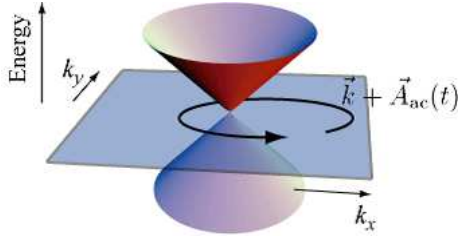}
\caption{
A trajectory of $\Vect{k}+\Vect{A}_{\rm ac}(t)$ around 
a Dirac point in a circularly-polarized light field. 
}
\label{fig1}
\end{figure}

Recently, a novel, optically controlled non-linear phenomenon 
has been proposed in ref. \cite{Okaprb}, 
in which the present authors 
revealed a new type of  Hall effect --- photovoltaic Hall effect --- 
can emerge in honeycomb lattices such as graphene subject to intense, circularly-polarized light 
despite the absence of static magnetic fields. 
If we denote the gauge field representing 
the circularly-polarized light as $\Vect{A}_{\rm ac}(t)=
\frac{F}{\Omega}(\cos\Omega t,\sin\Omega t)$,
where $F=eE$ is the field strength and $\Omega $ the 
frequency of the light, the 
tight-binding Hamiltonian for electrons reads
\begin{eqnarray}
H(t)=-\sum_{ij}t_{ij}e^{-i\hat{e}_{ij}
\cdot{\bf A}_{\rm ac}(t)}c_{i}^\dagger c_j ,
\end{eqnarray}
where $t_{ij}=w$ is the nearest-neighbor 
hopping for the honeycomb lattice, $\hat{e}_{ij}$ the 
unit vector connecting the bond between $i$ and $j$, and 
spins are ignored.  
The Hamiltonian is time periodic, and, in momentum space the $k$-points 
are driven in the Brillouin zone by the 
field as $\Vect{k}(t)=\Vect{k}-\Vect{A}_{\rm ac}(t)$.  
The orbit is a circle centered at $\Vect{k}$
with a radius $F/\Omega$, and the time period is 
$T=2\pi/\Omega$ (Fig.~\ref{fig1}). 

During the motion, the electron wave function 
acquires a nontrivial geometric phase, 
known as the Aharnov-Anandan phase\cite{Aharonov1987} 
which is the 
non-adiabatic extension of the Berry phase\cite{Berry1984} 
and reduces to the Berry phase in the 
adiabatic limit $\Omega\to 0$ with a fixed $F/\Omega$.  
An important feature in the electron wavefunctions 
in time-periodic systems, 
which is a temporal analogue of Bloch's theorem for 
spatially periodic systems, 
 is that the solution 
 $|\Psi_\alpha(t)\ket$ of the
time-dependent Schr\"odinger 
equation can be 
described by the time-periodic
Floquet states $|\Phi_\alpha(t)\ket$
as  $|\Psi_\alpha(t)\ket=e^{-i\ve_\alpha t}|\Phi_\alpha(t)\ket$,
where $\ve_\alpha$ is the Floquet quasi-energy 
and $\alpha$ an index that labels the Floquet states.
Note that $\alpha$ becomes a 
composite index $\alpha=(i,m)$ when the 
original system has an internal degrees of freedom $i$, e.g. 
the band index for multiband systems such as the 
Dirac cone with electron and hole branches, while 
$m=0,\pm 1,\pm 2,\ldots$ denotes the number of 
absorbed photons. 

The applied intense laser induces nontrivial photovoltaic transports as 
follows:  
Note first that we have two external fields, the intense laser 
irradiation 
and (ii) a weak, dc electric field for measuring the Hall effect.  
If we define the inner product averaged over a period of 
the ac laser field by
$\dbra\alpha|\beta\dket\equiv \frac{1}{T}\int_0^T dt\bra \alpha(t)|
\beta(t)\ket$,
the Floquet states span a complete orthogonal 
basis in the presence of a strong laser field, and
one can perform a perturbation expansion in the 
small dc field to probe the (Hall) conductivity 
to obtain a Kubo formula extended to the case of 
strong ac irradiation \cite{Okaprb},
\begin{eqnarray}
&&\sigma_{ab}(\Vect{A}_{ac})=i\int \frac{d\Vect{k}}{(2\pi)^d}
\sum_{\alpha,\beta\ne\alpha}
\frac{[f_\beta(\Vect{k})-f_\alpha(\Vect{k})]}{\ve_\beta(\Vect{k})-\ve_\alpha(\Vect{k})}
\frac{
\dbra\Phi_\alpha(\Vect{k})|J_b|\Phi_{\beta}(\Vect{k})\dket
\dbra\Phi_\beta(\Vect{k})|J_a|\Phi_{\alpha}(\Vect{k})\dket
}{\ve_\beta(\Vect{k})-\ve_\alpha(\Vect{k})+i\eta},
\end{eqnarray}
where $f_\alpha(\Vect{k})$ is the non-equilibrium distribution 
(occupation fraction) of the $\alpha$-th Floquet state, 
$\Vect{J}$ the current operator, and $\eta$ a positive infinitesimal.
The difference from the conventional 
Kubo formula in the absence of AC fields is that 
the energy is replaced with the Floquet quasi-energy, and 
the inner product with a time average. 
The photovoltaic Hall conductivity can be further simplified 
to \cite{Okaprb},  
\begin{eqnarray}
\sigma_{xy}(\Vect{A}_{\rm ac})=e^2\int \frac{d\Vect{k}}{(2\pi)^d}\sum_\alpha
f_\alpha(\Vect{k})\left[\nabla_{\Vect{k}}\times\Vect{\mathcal{A}}_\alpha(\Vect{k})\right]_z ,
\label{eq:TKNN}
\end{eqnarray}
in terms of a gauge field 
$\Vect{\mathcal{A}}_\alpha(\Vect{k}) \equiv 
-i\dbra\Phi_\alpha(\Vect{k})|\nabla_{\Vect{k}}|\Phi_\alpha(\Vect{k})\dket$.
This reduces to the TKNN formula \cite{TKNN}
in the adiabatic limit.
Note that the photovoltaic Hall effect does not take 
place in a linearly polarized light \cite{syzranov:045407}
which does not break the time-reversal symmetry.

\begin{figure}[tbh]
\centering 
\includegraphics[width=12.5cm]{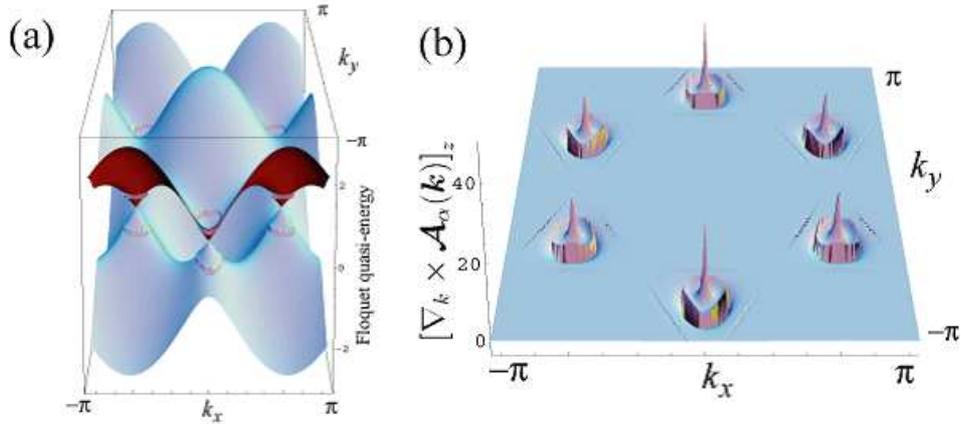}
\caption{
(a) The Floquet quasi-energy $\ve$ against the wave number, and 
(b) the photovoltaic Berry curvature $
\left[\nabla_k\times\Vect{\mathcal{A}}_\alpha(\Vect{k})\right]_z$
for the upper band for $F=0.1w,\;\Omega=1.0w$ in the honeycomb lattice. 
}
\label{fig2}
\end{figure}

\section{Photovoltaic Berry curvature in the honeycomb lattice}
Let us have a closer look at the Floquet states 
and the photovoltaic curvature $\left[\nabla_{\Vect{k}}\times\Vect{\mathcal{A}}_\alpha(\Vect{k})\right]_z $
in the honeycomb lattice.
We begin by noting that there exists infinite array of photo-dressed states
in the Floquet spectrum, whose quasi-energy is 
shifted by $m\Omega$ ($m=0,\pm 1,\ldots $)
from the plotted ones.
In Fig.~\ref{fig2}~(a), we plot the Floquet quasi-energy 
against wave number for the Floquet states having the 
largest weight of the $m=0$ Floquet state 
(a descendant of the original $F=0$ dispersion).
An important feature in the plot is that the 
band-gap-like
structure around $\ve\sim \pm \Omega/2$, 
where the gap opens an outcome of the photo-induced hybridization 
of levels, a first-order effect in $F$. 
More importantly, a small 
gap opens \cite{Okaprb} at the Dirac points
around $\ve\sim 0$, which is a consequence of 
circularly-polarized light.

The photovoltaic Berry curvature is plotted in
Fig.~\ref{fig2}~(b) for the same parameters.
Around each K-point we find 
a peak surrounded by concentric circles
and lines. The peak is due to the 
gap opening at the Dirac points, 
we can see that the sign is the 
same for both $K$ and $K'$ points.
They contribute equally to the photovoltaic Hall coefficient
in the momentum integral 
in eqn.~(\ref{eq:TKNN}) and {\it no cancellation 
between $K$ and $K'$ points occurs}.  
On the other hand, the concentric circles
and lines are due to the photo-induced hybridization,
e.g.,  the former corresponds to the band 
structure around
$\ve\sim \pm \Omega/2$ in Fig.~\ref{fig2}~(a).

\section{Experimental feasibility}
The two major candidates for the observation of the 
photovoltaic Hall effect are
\begin{itemize}
\item Graphene, including multi-layer systems,
\item Surface states in topological insulators,
\end{itemize}
in which Dirac-like dispersions are realized.
As for the experimental feasibility, 
a typical intensity of laser conceived here, 
$F\sim 0.001w$, corresponds to 
$E\sim 10^7\mbox{V/m}$ 
for photon energy $\Omega\sim 1\mbox{eV}$, $w=2.7\mbox{eV},\;a=2.6\AA$.
The size of the photovoltaic Dirac gap $\Delta=2\kappa$
 scales as
\begin{eqnarray}
\Delta \propto F^2/\Omega^3
\end{eqnarray}
for small field strengths, which was derived in ref.~\cite{Okaprb}.  
This formula implies that with a smaller laser energy $\Omega$ 
we can realize the photovoltaic Hall effect with a 
weaker laser strength $F$. 
Hence it should be interesting to study the 
$\Omega$ dependence of the photovoltaic Hall
effect. Another important point is that
in graphene, the Dirac dispersion is 
realized near zero energy, but the dispersion 
begins to reflect the lattice structure for higher energies, 
hence for higher values of $\Omega$.  
So we may ask: Will the Hall effect survive when $\Omega$ 
becomes large, say 
comparable with the band width?

\begin{figure}[t]
\centering 
\includegraphics[width=16.cm]{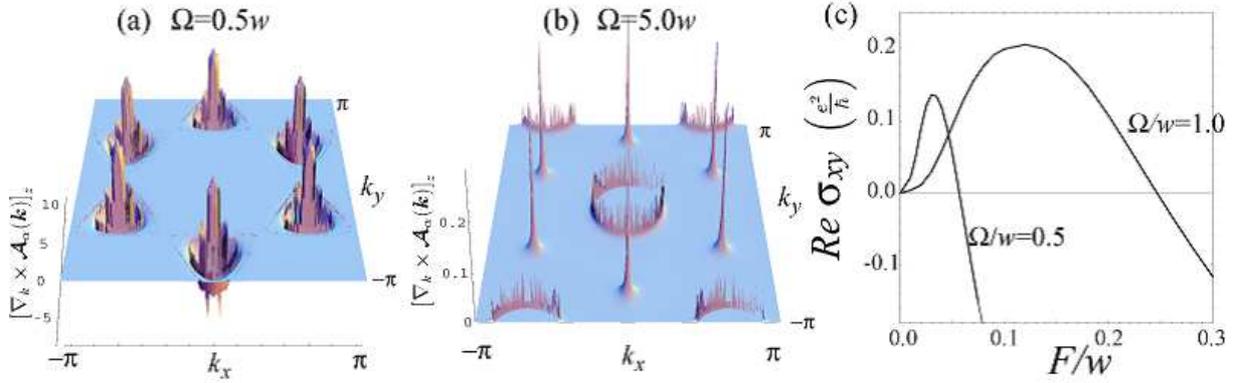}
\caption{
Photovoltaic Berry curvature 
for (a) $\Omega=0.5w$
and (b) $\Omega=5.0w$
with the same field strength as Fig.~\ref{fig2}. 
(c) The dependence of the
Photovoltaic Hall coefficient $\mbox{Re}~\sigma_{xy}$ on the 
strength of the ac-electric field $F$ in the honeycomb lattice. 
}
\label{fig3}
\end{figure}

In order to clarify this point, we have 
calculated the photovoltaic Berry curvature for 
several values of $\Omega$ in Fig.~\ref{fig3}.
For a smaller frequency in Fig.~\ref{fig3}~(a), 
the central peak becomes smaller and broader.
The surrounding concentric circles
become closely packed and increase in number 
 because we now have contributions from 
 photo-induced hybridization 
around $\ve\sim \pm \Omega$
in addition to $\ve\sim\pm\Omega/2$.  
For a larger frequency in Fig.~\ref{fig3}~(b), on the other hand, 
the peaks survive 
even though $\Omega$ is greater than the 
band width.  
However, the peak becomes smaller (and 
narrower), so it will become more difficult to observe the effect.

 In Fig.~\ref{fig3}~(c), we plot the 
Photovoltaic Hall coefficient (eqn.(3)) 
against the strength of the circularly polarized light.
Here, we assumed
sudden switch on of the electric fields, which implies
$f_\alpha=\sum_{E_n<E_F}|\dbra \Phi_\alpha|\psi_n\dket|^2$, 
where $|\psi_n\ket$ is the zero field eigen-wavefunction, and $E_n$ 
its energy. This 
assumption is expected to breakdown in the presence of
dissipation, however, the basic trend 
do not change which was confirmed by calculations
based on more elaborate techniques \cite{Okaprb}.
As expected from the scaling relation
of the photovoltaic Dirac gap $\Delta$ (eqn.~(4)),
the Hall coefficient first increases as $
\mbox{Re}~\sigma_{xy}\propto \Delta\propto F^2/\Omega^3$ 
in the weak field limit. At this field strength, 
the physics can be understood by the central peaks
of the Berry curvature at the K and K' points.
However, after saturating at a
certain value, the Hall coefficient starts to decrease and become 
negative. When this happens, the electrons
are deeply in the ac-Wannier Stark ladder regime 
and are localized
due to the strong circularly polarized light. 
The Berry curvature peaks around the K and K' points 
now become highly complex and the sign
alters at each concentric circles 
as shown in Fig.~\ref{fig3}~(a). This leads to negative
Hall coefficients after integrating over the Brillouin Zone.

\section{Conclusion}
We have studied the dependence of the photovoltaic Hall 
effect in a honeycomb lattice.  Especially, we have elaborated 
the dependence of the photovoltaic Berry curvature 
on the frequency $\Omega$ of the applied circularly-polarized light.
We have found that the photovoltaic Berry curvature
and thus, the Hall coefficient  
gradually increases as $\Omega$ becomes smaller
reflecting the relation $
\mbox{Re}~\sigma_{xy}\propto \Delta\propto F^2/\Omega^3$. 
However, even when $\Omega$ is large and the 
Dirac band approximation breaks down, 
we found evidence that the photovoltaic Hall 
effect survives, which indicates that
this is not limited to Dirac bands 
but the effect is more universal and can 
be realized in a wide range of electron systems in 
circularly polarized light.


TO was supported by Grant-in-Aid for young Scientists (B).



\end{document}